\documentstyle[preprint,aps]{revtex}
\tighten
\begin{document}
\preprint{Preprint Numbers:
\parbox[t]{50mm}{ADP-94-15/T156\\
		 FSU-SCRI-94-98 }}
\draft
%_______________________ Title, Authors ____________________________________
\title{Chiral Symmetry Breaking in Quenched Massive
       Strong-Coupling QED$_4$}
\author{
  Frederick T. Hawes\footnotemark[1]
	  and
  Anthony G. Williams\footnotemark[2]
  \vspace*{2mm} }

\address{
  \footnotemark[1] Department of Physics and SCRI,
    Florida State University, \\
    Tallahassee, Florida 32306-3016
  \vspace*{2mm}\\
  \footnotemark[2] Department of Physics and Mathematical Physics, \\
    University of Adelaide, \\
    South Australia 5005, Australia
}
%\date{}
%
\maketitle
%%-------------------------------------------------------------------%
%
\begin{abstract}
  We present results from a study of subtractive renormalization
  of the fermion propagator
  Dyson-Schwinger equation (DSE) in massive strong-coupling quenched
  QED$_4$.  Results are compared for three different fermion-photon
  proper vertex
  {\it Ans\"{a}tze\/}: bare $\gamma^\mu$, minimal Ball-Chiu,
  and Curtis-Pennington.
  The procedure is straightforward to implement and numerically stable.
  This is the first study in which this technique is used and
  it should prove useful in future DSE studies, whenever renormalization
  is required in numerical work.
\end{abstract}

\section{Introduction}

Strong coupling QED in three space and one time dimension has been
studied within the Dyson-Schwinger Equation (DSE) formalism for some
time \cite{BJW,FGMS,Mandula},
in order to see whether there may be a phase transition to a
nontrivial ``local'' theory at high momenta
\cite{FGMS,Miransk1,Miransk2,Miransk3},
as a model for dynamical chiral symmetry breaking (DCSB) in walking
technicolor theories \cite{techni,Mahanta},
and also as an abelianized model for nonperturbative phenomena
in QCD \cite{AbQCD,AtkJ}.  For a recent review of
Dyson-Schwinger equations and their application see Ref.~\cite{TheReview}.
The usual approach is to write the DSE for the fermion propagator or
self-energy, possibly including equations for the photon vacuum
polarization \cite{Rakow,Pi-also} or
the fermion-photon proper vertex \cite{BJW}.
An {\it Ansatz\/} is made for the undefined Green's
functions that contain the infinite continuation of the tower
of DSE's.  The resulting nonlinear integral equations are solved
numerically in Euclidean space \cite{GCW} by iteration.
DCSB occurs when the fermion propagator develops a nonzero scalar
self-energy in the absence of an explicit chiral symmetry breaking
(ECSB) fermion mass.  We refer to coupling constants strong enough to induce
DCSB as supercritical and those weaker are called subcritical.
We write the fermion propagator as
\begin{equation}
  S(p) = \frac{Z(p^2)}{\not\!p - M(p^2)}
       = \frac{1}{A(p^2) \not\!p - B(p^2)}
\end{equation}
with $Z(p^2)$ the finite momentum-dependent fermion renormalization,
and $B(p^2)$ the scalar self-energy.  In the massless theory
(i.e., in the absence of an ECSB bare electron mass) by definition
DCSB occurs when $B(p^2)\neq 0$.
Note that $A(p^2)\equiv 1/Z(p^2)$ and $M(p^2)\equiv B(p^2)/A(p^2)$.

%%Many of these studies, even until quite recently, have used the
Many studies, even until quite recently, have used the
bare vertex as an {\it Ansatz\/} for the one-particle
irreducible (1-PI) vertex $\Gamma^\nu(k,p)$
\cite{FGMS,Miransk1,Miransk2,Mahanta,AbQCD,Rakow,Pi-also,Bare},
despite the fact that this violates the Ward-Takahashi Identity (WTI)
\cite{WTI}.  It is also common, especially in studies motivated by
walking technicolor theories \cite{techni}, to find vertex
{\it Ans\"{a}tze\/} which claim to solve the WTI, but which still
possess kinematic singularities in the limit of zero photon momentum
$q^2 = (k-p)^2 \to 0$ \cite{KKM,bad_Gamma}.
With any of these {\it Ans\"{a}tze\/} the resulting fermion propagator
is not gauge-covariant, i.e., physical quantities such as the critical
coupling for dynamical symmetry breaking, or the mass itself, are
gauge-dependent \cite{KKM}.
A general form for $\Gamma^\nu(k,p)$ which does satisfy the Ward
Identity was given by Ball and Chiu in 1980 \cite{BC}; it consists of a
minimal longitudinally constrained term which satisfies the WTI, and a set
of tensors
spanning the subspace transverse to the photon momentum $q$.

Although the WTI is necessary for gauge-invariance, it is not a
sufficient condition; further, with many of these vertex
{\it Ans\"{a}tze\/} the fermion propagator DSE is not multiplicatively
renormalizable, which is equivalent to saying that
overlapping logarithms are present.
There has been much recent research on the use
of the transverse parts of the vertex to ensure both gauge-covariant
and multiplicatively renormalizable solutions
\cite{King,gaugetech,HaeriQED,CPI,CPII,CPIII,CPIV,dongroberts,BashPenn},
some of which will be discussed below.

What is common to essentially all of the studies so far is
that the fermion propagator is not in practice subtractively renormalized.
Most of these studies have assumed an initially massless theory
and have renormalized at the ultraviolet cutoff of the integrations,
taking $Z_1 = Z_2 = 1$.  Where a nonzero bare mass has been used
\cite{Miransk3,Rakow}, it has simply been added to the scalar term in the
propagator.  Although there have been formal discussions
of the renormalization \cite{Miransk2,Miransk3,CPI,CPIV},
the important step of subtractive renormalization has not been performed.

We describe here the results of a study of subtractive
renormalization in the fermion DSE, in quenched strong-coupling
QED$_4$.  (In the context of this study of QED, the term ``quenched'' means
that the bare photon propagator is used in the fermion self-energy DSE,
so that $Z_3 = 1$.  Virtual fermion loops may still be present, however,
within the vertex corrections.)  We believe this is the first
calculation in which the subtraction is performed numerically
as the integral equations are being iterated.
This new technique should prove useful in other DSE studies.
Results are obtained for DSE with three different vertices:
the bare $\gamma^\mu$, the minimal Ball-Chiu vertex form \cite{BC},
and the Curtis-Pennington vertex \cite{CPI,CPII,CPIII,CPIV}.

Organization of the paper is as follows.  The DSE for the renormalized
fermion propagator, and {\it Ans\"{a}tze\/} for the proper vertex,
are discussed in section \ref{sec_method}.
The subtractive renormalization is described in section
\ref{sec_subtr}.
Section \ref{sec_subcritical} presents results at subcritical couplings,
while Section \ref{sec_DCSB} discusses the results for couplings above
the DCSB threshold.  Conclusions and discussion are given in section V.

\section{DSE and Vertex Ans\"{a}tze}
\label{sec_method}

The DSE for the renormalized fermion propagator, in a general covariant
gauge, is
\begin{equation} \label{fermDSE_eq}
  S^{-1}(p^2) = Z_2(\mu,\Lambda)[\not\!p - m_0(\Lambda)]
    - i Z_1(\mu,\Lambda) e^2 \int^{\Lambda} \frac{d^4k}{(2\pi)^4}
	  \gamma^{\mu} S(k) \Gamma^{\nu}(k,p) D_{\mu \nu}(q)\:;
\end{equation}
here $q=k-p$ is the photon momentum, $\mu$ is the renormalization
point, and $\Lambda$ is a regularizing parameter (taken here to be an
ultraviolet momentum cutoff).  We write
$m_0(\Lambda)$ for the regularization-parameter dependent bare mass.
In the massless theory (i.e., in the absence of an ECSB) the bare
mass is zero, $m_0(\Lambda)=0$.
The physical charge is $e$ (as opposed to the bare charge $e_0$),
and the general form for the photon propagator is
\begin{equation}
  D^{\mu\nu}(q) = \left\{
    \left( -g^{\mu\nu} + \frac{q^\mu q^\nu}{q^2} \right)
    \frac{1}{1-\Pi(q^2)} - \xi \frac{q^\mu q^\nu}{q^2} \right\}\:,
\end{equation}
with $\xi$ the covariant gauge parameter.  Since we will work in the
quenched approximation and the Landau gauge we have
$e^2 \equiv e_0^2 = 4\pi\alpha_0$ and
\begin{equation}
  D^{\mu\nu}(q) \to  D_0^{\mu\nu}(q)
    = \left( -g^{\mu\nu} + \frac{q^\mu q^\nu}{q^2} \right)
          \frac{1}{q^2}\:,
\end{equation}
%%\begin{eqnarray}
%%  D^{\mu\nu}(q) & \equiv & D_0^{\mu\nu}(q) \nonumber\\
%%  & = & \left\{ \left( -g^{\mu\nu} + \frac{q^\mu q^\nu}{q^2} \right)
%%          - \xi \frac{q^\mu q^\nu}{q^2} \right\}\frac{1}{q^2}\:.
%%\end{eqnarray}
for the photon propagator.

\subsection{Vertex Ansatz}
\label{subsec_QEDvtx}

The requirement of gauge invariance in QED leads to the Ward-Takahashi
Identities (WTI); the WTI for the fermion-photon vertex is
\begin{equation}
  q_\mu \Gamma^\mu(k,p) = S^{-1}(k) - S^{-1}(p)\;,
\end{equation}
where $q = k - p$\/ \cite{BjD,MnS}.  This is a generalization of the
original differential Ward identity, which expresses the effect of
inserting a zero-momentum photon vertex into the fermion propagator,
\begin{equation}
  \frac{\partial S^{-1}(p)}{\partial p_\nu} = \Gamma^{\nu}(p,p)\:.
\end{equation}
In particular, it guarantees the equality of the propagator
and vertex renormalization constants, $Z_2 \equiv Z_1$.
The Ward-Takahashi Identity is easily shown to be satisfied
order-by-order in perturbation theory and can also be derived
nonperturbatively.

As discussed in \cite{TheReview,Craig_1}, this can be thought
of as just one of a set of six general requirements on the vertex:
(i) the vertex must satisfy the WTI; (ii) it should contain no kinematic
singularities; (iii) it should transform under charge conjugation ($C$),
parity inversion ($P$), and time reversal ($T$) in the same way
as the bare vertex, e.g.,
	\begin{equation}
	  C^{-1} \Gamma_\mu(k,p) C = - \Gamma_\mu^{\sf T}(-p,-k)
	\end{equation}
(where the superscript {\sf T} indicates the transpose);
(iv) it should reduce to the bare vertex in the weak-coupling
limit; (v) it should ensure multiplicative renormalizability of the
DSE in Eq. (\ref{fermDSE_eq});
(vi) the transverse part of the vertex should be specified to
	ensure gauge-covariance of the DSE.

Ball and Chiu \cite{BC} have given a description of the most general
fermion-photon vertex that satisfies the WTI; it consists of a
longitudinally-constrained (i.e., ``Ball-Chiu'') part
$\Gamma^\mu_{\rm BC}$, which is a minimal solution of the WTI,
and a basis set of eight transverse vectors $T_i^\mu(k,p)$,
which span the hyperplane specified by $q_\mu T_i^\mu(k,p) = 0$,
$q \equiv k-p$.
The minimal longitudinally constrained part of the vertex is given by
\begin{equation} \label{minBCvert_eqn}
  \Gamma^\mu_{\rm BC}(k,p) = \frac{1}{2}[A(k^2) +A(p^2)] \gamma^\mu
    + \frac{(k+p)^\mu}{k^2-p^2}
      \left\{ [A(k^2) - A(p^2)] \frac{{\not\!k}+ {\not\!p}}{2}
	      - [B(k^2) - B(p^2)] \right\}\:.
\end{equation}
The transverse tensors are given by
\begin{eqnarray}
  T_1^\mu(k,p) & = & p^\mu (k \cdot q) - k^\mu (p \cdot q)\:,
                                                        \label{T1mu}\\
  T_2^\mu(k,p) & = & ({\not\!p} + {\not\!k}) T_1^\mu\:,     \label{T2mu}\\
  T_3^\mu(k,p) & = & q^2 \gamma^\mu - q^\mu \not\!q\:,  \label{T3mu}\\
  T_4^\mu(k,p) & = & p^\nu k^\lambda \sigma_{\nu\lambda} T_1^\mu\:,
							\label{T4mu}\\
  T_5^\mu(k,p) & = & \sigma^{\mu\nu} q_\nu\:,           \label{T5mu}\\
  T_6^\mu(k,p) & = & \gamma^\mu (k^2-p^2) - (k+p)^\mu \not\!q\:,
                                                        \label{T6mu}\\
  T_7^\mu(k,p) & = & \frac{(k^2-p^2)}{2}
    [\gamma^\mu ({\not\!p} + {\not\!k}) - (k+p)^\mu ]
    + (k+p)^\mu p^\nu k^\lambda \sigma_{\nu\lambda}\:,  \label{T7mu}\\
  T_8^\mu(k,p) & = & -\gamma^\mu p^\lambda k^\nu \sigma_{\lambda\nu}
      - p^\mu\not\!k - k^\mu\not\!p\:.                \label{T8mu}
\end{eqnarray}
A general vertex is then written as
\begin{equation} \label{anyfullG_eqn}
  \Gamma^\mu(k,p) = \Gamma_{BC}^\mu(k,p)
    + \sum_{i=1}^{8} \tau_i(k^2,p^2,q^2) T_i^\mu(k,p)\:,
\end{equation}
where the $\tau_i$ are functions which must be chosen to give the
correct $C$, $P$, and $T$ invariance properties.

The work of Curtis and Pennington \cite{CPI,CPII,CPIII,CPIV} was
mentioned above in connection with the specification of the transverse
vertex terms in order to produce gauge-invariant and multiplicatively
renormalizable solutions to the DSE.
In the framework of massless QED$_4$, they eliminate
four of the transverse vectors since they are Dirac-even and must
generate a scalar term.  By requiring that the vertex $\Gamma^\mu(k,p)$
reduce to the leading log result for $k \gg p$ they are led to
eliminate all the transverse basis vectors except $T_6^\mu$, with a
dynamic coefficient chosen to make the DSE multiplicatively
renormalizable.  This coefficient has the form
\begin{equation}
  \tau_6(k^2,p^2,q^2) = \frac{1}{2}[A(k^2) - A(p^2)] / d(k,p)\:,
\label{CPgamma1}
\end{equation}
where $d(k,p)$ is a symmetric, singularity-free function of $k$ and $p$,
with the limiting behavior $\lim_{k^2 \gg p^2} d(k,p) = k^2$.
[Here, $A(p^2)\equiv 1/Z(p^2)$ is their $1/{\cal F}(p^2)$.]
For purely massless QED, they find a suitable form,
$d(k,p) = (k^2 - p^2)^2/(k^2+p^2)$.  This is generalized to the
case with a dynamical mass $M(p^2)$, to give
\begin{equation}
d(k,p) = \frac{(k^2 - p^2)^2 + [M^2(k^2) + M^2(p^2)]^2}{k^2+p^2}\:.
\label{CPgamma2}
\end{equation}
They establish that multiplicative renormalizability is retained
up to next-to-leading-log order in the DCSB case.  Subsequent
papers establish the form of the solutions for the renormalization
and the mass \cite{CPIII} and studied the gauge-dependence
of the solutions \cite{CPIV}.  Dong, Munczek and Roberts
\cite{dongroberts} subsequently showed that the lack of
exact gauge-covariance of the solutions is in part due to the use of a hard
cutoff in the integral equations.
The fact that there is still some residual gauge-dependence in the
physical observables such as the chiral critical point shows that
the C-P vertex {\it Ansatz\/} is not yet the ideal choice.  Dong,
Munczek and Roberts \cite{dongroberts} derived an {\it Ansatz\/} for
the transverse vertex terms which satisfies the WTI and makes the
fermion propagator gauge-covariant.

Bashir and Pennington \cite{BashPenn} have recently described a
vertex {\it Ansatz\/} which makes the fermion self-energy exactly
gauge-covariant, in the sense that the critical point for the chiral
phase transition is independent of gauge.  Their work is
a continuation of that of Dong, Munczek and Roberts, and indeed their
vertex {\it Ansatz\/} corresponds to the general form suggested in
\cite{dongroberts}.  However, the kinematic factors $\tau_{2,3,6,8}$
are extremely complicated in form and depend upon a pair of as yet
undetermined functions $W_{1,2}(k^2,p^2)$ which must be chosen to
guarantee that the weak-coupling limit of $\Gamma^\mu$ matches the
perturbative result.  Renormalization studies of the DSE using this
vertex {\it Ansatz\/} should be interesting and represent a direction
for further research.

In any case, for the solutions to the fermion DSE using the C-P vertex, the
critical
point for the chiral phase transition has been shown to have a much
weaker gauge-dependence than that for the DSE with the bare or minimal
Ball-Chiu vertices \cite{ABGPR}.
In this work we will primarily use the Curtis-Pennington
{\it Ansatz\/} but for the sake of comparison
will also present results for the bare vertex.
Some solutions are also obtained with the minimal Ball-Chiu vertex
which, like the Curtis-Pennington vertex, satisfies the WTI, but
which does not lead to approximately gauge-invariant solutions nor
multiplicative renormalizability.

For each vertex {\it Ansatz\/}, the equations are separated into
a Dirac-odd part describing the finite propagator renormalization
$A(p^2)$, and a Dirac-even part for the scalar self-energy, by taking
$\frac{1}{4}{\rm Tr}$ of the DSE multiplied by $\not\!p/p^2$ and 1,
respectively.  The equations are solved in Euclidean space and so
the volume integrals
$\int d^4k$ can be separated into angle integrals and an integral
$\int dk^2$; the angle integrals are easy to perform analytically,
yielding the two equations which which will be solved numerically.

\section{The Subtractive Renormalization}
\label{sec_subtr}

The subtractive renormalization of the fermion propagator DSE proceeds
similarly to the one-loop renormalization of the propagator in QED.
(This is discussed in \cite{TheReview} and in \cite{IZ}, p.~425ff.)
One first determines a finite, {\it regularized\/} self-energy,
which depends on both a regularization parameter and the
renormalization point;
then one performs a subtraction at the renormalization point,
in order to define the renormalization parameters $Z_1$, $Z_2$, $Z_3$
which give the full (renormalized) theory in terms of the regularized
calculation.

A review of the literature of DSEs in QED shows, however, that this
step is never actually performed.  Curtis and Pennington \cite{CPIV}
for example, define their renormalization point at the UV cutoff.
Miransky \cite{Miransk2} gives a formal discussion of the variation
of the mass renormalization $Z_m(\mu,\Lambda)$, but does not implement
it numerically.

Many studies take $Z_1 = Z_2 = 1$
\cite{King,HaeriQED,CPI,CPII,CPIII,CPIV};
this is a reasonable approximation in cases where the coupling
$\alpha_0$ is small (i.e.,
$\alpha_0$ \raisebox{-.4ex}{$\stackrel{<}{\sim}$} 1.15),
but if $\alpha_0$ is chosen large enough,
the value of the dynamical mass at the renormalization point may be
significantly large compared with its maximum in the infrared.
For instance, in \cite{CPIV}, figures for the fermion mass are given
with $\alpha_0$ = 0.97, 1.00, 1.15 and 2.00 in various gauges.  For
$\alpha_0 = 2.00$, the fermion mass at the cutoff is down by only an
order of magnitude from its limiting value in the infrared.

Repeating the arguments from \cite{TheReview}, one defines
a regularized self-energy $\Sigma'(\mu,\Lambda; p)$, leading to the
DSE for the renormalized fermion propagator,
\begin{eqnarray} \label{ren_DSE}
  \widetilde{S}^{-1}(p) & = & Z_2(\mu,\Lambda) [\not\!p - m_0(\Lambda)]
    - \Sigma'(\mu,\Lambda; p) \nonumber\\
    & = & \not\!p - m(\mu) - \widetilde{\Sigma}(\mu;p)\:,
\end{eqnarray}
where the (regularized) self-energy is
\begin{equation} \label{reg_Sigma}
  \Sigma'(\mu,\Lambda; p) = i Z_1(\mu,\Lambda) e^2 \int^{\Lambda}
    \frac{d^4k}{(2\pi)^4} \gamma^\lambda \widetilde{S}(\mu;k)
      \widetilde{\Gamma}^\nu(\mu; k,p)
      \widetilde{D}_{\lambda \nu}(\mu; (p-k))\:.
\end{equation}
[To avoid confusion we will follow Ref.~\cite{TheReview} and in this
section {\it only} we will denote
regularized quantities with a prime
and renormalized ones with a tilde, e.g. $\Sigma'(\mu,\Lambda; p)$
is the regularized self-energy depending on both the renormalization
point $\mu$ and regularization parameter $\Lambda$
and $\widetilde{\Sigma}(\mu;p)$ is the renormalized self-energy.]
As suggested by the notation (i.e., the omission of the $\Lambda$-dependence)
renormalized quantities must become independent of the regularization-parameter
as the regularization is removed (i.e., as $\Lambda\to\infty$).
The self-energies are decomposed into Dirac and scalar parts,
\begin{displaymath}
  \Sigma'(\mu,\Lambda; p) = \Sigma'_d(\mu,\Lambda; p^2) \not\!p
		     + \Sigma'_s(\mu,\Lambda; p^2)
\end{displaymath}
(and similarly for the renormalized quantity,
$\widetilde{\Sigma}(\mu,p)$).
By imposing the renormalization boundary condition,
\begin{equation}
  \left. \widetilde{S}^{-1}(p) \right|_{p^2 = \mu^2}
  = \not\!p - m(\mu)\:,
\end{equation}
one gets the relations
\begin{equation}
  \widetilde{\Sigma}_{d,s}(\mu; p^2) =
    \Sigma'_{d,s}(\mu,\Lambda; p^2) - \Sigma'_{d,s}(\mu,\Lambda; \mu^2)
\end{equation}
for the self-energy,
\begin{equation}
  Z_2(\mu,\Lambda) = 1 + \Sigma'_d(\mu,\Lambda; \mu^2)
\end{equation}
for the renormalization, and
\begin{equation}
  m_0(\Lambda) = \left[ m(\mu) - \Sigma'_s(\mu,\Lambda; \mu^2) \right]
	/ Z_2(\mu,\Lambda)
\label{baremass}
\end{equation}
for the bare mass.  In order to reproduce the case with no ECSB mass,
for a given cutoff $\Lambda$, one chooses
$m(\mu) = \Sigma'_s(\mu,\Lambda; \mu^2)$
so that the bare mass $m_0(\Lambda)$ is zero.  The mass renormalization
constant is given by
\begin{equation}
  Z_m(\mu,\Lambda) = m_0(\Lambda)/m(\mu)\:,
\label{Z_m}
\end{equation}
i.e., as the ratio of the bare to renormalized mass.

The vertex renormalization, $Z_1(\mu,\Lambda)$ is identical to
$Z_2(\mu,\Lambda)$ as long as the vertex {\it Ansatz\/} satisfies
the Ward Identity; this is how it is recovered for multiplication
into $\Sigma'(\mu,\Lambda;p)$ in Eq. (\ref{reg_Sigma}).  It will be
noticed that this is inappropriate for the bare-vertex {\it Ansatz\/}
since it fails to satisfy the WTI; nonetheless, since for the bare vertex
case there is no way to
determine $Z_1(\mu,\Lambda)$ independently we will use $Z_1 = Z_2$ for the
sake of comparison.  In the Landau gauge for the bare
vertex these will then both be 1, since in this case
$\Sigma'_d(\mu,\Lambda; p^2) = 0$ for all $p^2$ as is well known
\cite{TheReview}.

We carry out this step of the renormalization, using the bare vertex
$\gamma^\mu$, the minimal Ball-Chiu vertex of Eq.~(\ref{minBCvert_eqn}),
and that with the Curtis-Pennington transverse term added
[see Eqs.~(\ref{CPgamma1},\ref{CPgamma2})].
The equations are again first written in Minkowski space, and then
rotated to Euclidean space.  These are then solved by iteration
from an initial guess, with a wide range of cutoffs, and with
various renormalization points $\mu^2$ and renormalized masses
$m(\mu)$.

\section{Results}

Solutions were obtained for the DSE with the Curtis-Pennington and
bare vertices, for couplings $\alpha_0$ from 0.1 to 1.75;
solutions were also obtained for the minimal Ball-Chiu vertex, with
couplings $\alpha_0$ from 0.1 to 0.6 (for larger couplings the DSE with
this vertex was susceptible to numerical noise).
In Landau gauge, the critical coupling for the DSE with bare vertex
is $\alpha_c^{\rm bare} = \pi/3$; the critical coupling
for the Curtis-Pennington vertex is $\alpha_c^{\rm CP} = 0.933667$
\cite{ABGPR}, and that for the Ball-Chiu vertex is expected to be
close to these two values.  Since there are some qualitative differences
between the solutions with subcritical and supercritical couplings,
we shall describe these separately.

\subsection{Subcritical Couplings}
\label{sec_subcritical}

Solutions for the Curtis-Pennington and bare vertex
{\it Ans\"{a}tze\/} were obtained for values of the coupling
$\alpha_0$ from 0.1 to 0.9.  The program using the Ball-Chiu
vertex was also run, with couplings $\alpha_0$ from 0.1 to 0.6.
For the solutions in the subcritical range, the renormalization point
$\mu^2 = 100$, and renormalized masses were either $m(\mu) = 10$ or
30.  Ultraviolet cutoffs were $1.0\times 10^{12}$ and
$1.0\times10^{18}$.

The family of solutions for the Curtis-Pennington vertex
with $m(\mu) = 10$ is shown in Fig.~\ref{fig_family}.
Fig.~\ref{fig_family} a) shows the finite renormalization
$A(p^2)$; note that for the solutions with bare $\gamma^\nu$, we
would have $A(p^2) \equiv 1$ for all values of the coupling.
Fig.~\ref{fig_family} b) shows the masses $M(p^2)$.

Since the equations have no inherent mass-scale, the cutoff
$\Lambda$, renormalization point $\mu$, $m(\mu)$, and units of
$M(p^2)$ or $B(p^2)$ all scale multiplicatively, and the units are
arbitrary.  For instance, the solutions for any given coupling
in Fig.~\ref{fig_family} could represent a particle with mass
10MeV at the renormalization point $\mu^2 = -100 {\rm MeV}^2$,
with the units of $p^2$ in ${\rm MeV}^2$ and those of $B$ in MeV;
or it could represent one with units in GeV, or in electron masses.
Since in four dimensions the coupling has no mass dimension,
it would remain unchanged for all such choices of units.
This is true for all the graphs that follow.

A comparison of the solutions for the three different vertices, for
the coupling $\alpha_0 = 0.5$, is given in Fig.~\ref{fig_compare3}.
As can be seen in Fig.~\ref{fig_compare3} a), the finite
renormalization $A(p^2)$ deviates less from unity
in the Curtis-Pennington case.
The scalar self-energy curves $B(p^2)$, shown in Fig.~\ref{fig_compare3}
b), look qualitatively similar. However, since $M(p^2)=B(p^2)/A(p^2)$,
there are differences both in the
infrared value of mass (for which we have
$M_{\rm CP} < M_{\rm BC} < M_{\rm bare}$ for the same $\alpha_0$),
and in the asymptotic scaling.

The slight difference in scaling of the mass functions for different
vertex {\it Ans\"{a}tze\/} is shown in Fig.~\ref{scaled_Ms},
for $\alpha_0 = 0.5$.
Miransky has shown \cite{Miransk2} that in Landau gauge with the
rainbow approximation, the fermion mass scales asymptotically
as $M(p^2) \sim (p^2)^{-(\frac{1}{2}-\gamma')}$ with the anomalous
dimension
$\gamma'(\alpha_0) = \frac{1}{2} \sqrt{1 - \alpha_{0}/(\pi/3)}$.
For $\alpha_0 = 0.5$, the anomalous dimension is $\gamma' = .3614329$,
giving the power $\frac{1}{2} - \gamma' = 0.138567$.
When all the solutions are multiplied
by $(p^2)^{\frac{1}{2} -\gamma'(0.5)}$,
the difference in anomalous dimensions shows up
as a difference in slope of the mass curves on a log-log plot.
The anomalous dimensions for the mass functions
with Ball-Chiu and Curtis-Pennington vertices are found
graphically, to be approximately
$\gamma_{\rm BC} = 0.35836$ giving the power
$ \frac{1}{2} - \gamma_{\rm CP} = 0.14164$,
and $\gamma_{\rm CP} = 0.35843$ giving the power
$ \frac{1}{2} - \gamma_{\rm CP} = 0.14157$, respectively.  The small dips
apparent at the end of two of these curves are due to having a hard
cut-off $\Lambda^2=10^{12}$.  As $\Lambda$ is increased these move to
higher momenta also.

For the equations with the Curtis-Pennington vertex,
the renormalization constants $Z_2$ and $Z_m$,
and the bare mass $m_0(\Lambda)$ were evaluated for
$\alpha_0 = 0.1$ through 0.9,
using cutoffs $1 \times 10^{12}$ and $1 \times 10^{18}$
and renormalization conditions $\mu^2 = 100$, $m(\mu) = 10$.
These are presented in Table \ref{table_Zs_by_a}.
Specifically for $\alpha_0 = 0.5$,
the renormalization constants $Z_2(\mu,\Lambda)$ and $Z_m(\mu,\Lambda)$
were calculated for several values of $\mu$,
using cutoffs $\Lambda^2 = 1 \times 10^{12}$ and $1 \times 10^{18}$.
These are given in Table \ref{table_Zs},
and the mass renormalizations $Z_m$ are plotted
as a function of $\mu^2/\Lambda^2$
in Fig.~\ref{fig_Zm_scaling}.
All these solutions used bare masses chosen so as to duplicate the
mass solution for $\mu^2 = 100$, $m(\mu) = 10$.
It can be seen that as the value of $\mu$ increases, $Z_2$
approaches 1, and also that $Z_2$ is almost independent of the
cutoff $\Lambda$, at least for $\Lambda \gg \mu$.
The mass renormalization also scales asymptotically (i.e., for sufficiently
large $\Lambda$ and $\mu$) as
$Z_m(\mu,\Lambda) \propto
    (\mu^2/\Lambda^2)^{\frac{1}{2} - \gamma_{\rm CP}}$.
The slight deviation from this scaling behaviour evident in
Fig.~\ref{fig_Zm_scaling}
occurs in the region where $\mu$ is not sufficiently large.

The solutions are extremely stable as the cutoff is varied.
The ultraviolet momentum cutoff is varied by six orders of magnitude with
the mass unchanged within numerical accuracy for sufficiently large
$\Lambda$.

\subsection{Results above DCSB Threshold}
\label{sec_DCSB}

The program using the Curtis-Pennington vertex was run
with several values of the coupling $\alpha_0$
from 0.97 to 1.75, and with several different cutoffs.
In all cases the mass and finite renormalization were stable with respect
to very large variations in cutoff.  Typical results
for the self-energy $B(p^2)$ are given in Fig.~\ref{Ren_B}.
These were generated using
$\alpha_0 = 1.00$, $\mu^2 = 100$, and $m(\mu) = 10$ (in arbitrary units).
In both graphs, the cutoff is stepped by factors of 100 over eight
orders of magnitude, and again the variation in the solutions is of the
order of the tolerance of the calculation.

The bare-vertex DSE was run with couplings $\alpha_0 = 1.15$, 1.5
and 1.75 for comparison to the Curtis-Pennington solutions.  A
comparison of typical mass curves is given in Fig.~\ref{fig_CPvsbare}.
Here the mass functions are compared directly since for the bare
vertex the scalar self-energy is exactly the mass.

An unexpected feature of the equations (for supercritical couplings only)
is that for any set
$\{\alpha_0, \mu^2, m(\mu)\}$, as the cutoff is increased there
is a small region where the dynamical mass becomes negative.
This behavior
was verified to be insensitive to number of grid points, and
was stable as the cutoff was increased above the region of the
negative self-mass, so it seems to be a real effect.
The significance of this for QED is not completely understood.
One possibility is that it may signal the the failure of
multiplicative renormalizability for the model DSE, in which case
further refinements of the vertex {\it Ansatz\/} may be called for.
It should be emphasized that this negative dip in the scalar self-energy
is very small unless the coupling $\alpha_0$ approaches 2.
For instance, for $\alpha_0 = 1.00$, the negative peak was
$\sim 2.4 \times 10^{-5}$ the size of the renormalized mass.
Specifically, we found that for
$\mu^2 = 100$, $m(\mu) = 10$,
the values are $B(0) = 10.745$, $B(1.634 \times 10^{12}) = -.00024$;
for the case with $\alpha_0 = 1.75$,
$\mu^2 = 1 \times 10^4$, $m(\mu) = 500$,
the values are $B(0) = 531.8$, $B(3.338 \times 10^{8}) = -28.3$.
An example is shown for $\alpha_0=1.75$, $\mu^2 = 1 \times 10^4$,
$m(\mu) = 500$ in Fig.~\ref{fig_zcross}.

There is no apparent scaling behavior for the mass renormalization $Z_m$ in
the supercritical case.
The renormalization constant $Z_2(\mu,\Lambda)$ remains finite and
well-behaved with increasing $\Lambda$ as expected in the Landau gauge.
Typical values, using $\alpha_0 = 1.15$ and the Curtis-Pennington vertex,
are given in Table \ref{table_Zs_1.15}.
{}From the results in this table and from Eqs.~(\ref{baremass},\ref{Z_m})
we see that $\Sigma'_s(\mu,\Lambda;\mu)$ first increases beyond $m(\mu)$
as $\Lambda$ increases so that $m_0(\Lambda)$ goes negative. As $\Lambda$ is
further increased the last line of this table indicates that $\Sigma'_s$
then begins to decrease.

\section{Summary and Conclusions}

We have described preliminary results in a study of four-dimensional
quenched QED, with subtractive renormalization performed numerically,
during the calculation.  We believe that this is the
first calculation of its kind, and the technique described here will
be applicable elsewhere (e.g., also in QCD), whenever numerical
renormalization is required.  The importance of this approach
is that it allows studies using different regularization procedures
(e.g., various soft versus hard momentum cut-offs) and/or different
cut-offs to be meaningfully compared at the same renormalization point.
This includes also comparisons with results from lattice studies of QED,
which should prove useful in providing further guidance in the choice
of reasonable {\it Ans\"atze} for the vertex and photon propagator.
Without renormalization only the unrenormalized, regulated quantities
would be be obtained and any such comparisons would be meaningless.
In addition,
in order to study the nonperturbative behaviour of renormalization constants
such as $Z_1(\mu,\Lambda)$, $Z_2(\mu,\Lambda)$, and $Z_m(\mu,\Lambda)$ they
must be numerically extracted and so a method such at that described here
would be essential.

The Curtis-Pennington vertex has been the primary focus of this study,
since it has the desirable properties of making the solutions approximately
gauge-invariant and also multiplicatively renormalizable up to
next-to-leading log order.
Solutions have been obtained for comparison purposes, using the minimal
Ball-Chiu vertex and using the bare vertex {\it Ans\"{a}tze\/}
(with $Z_1=Z_2$).
For the Ball-Chiu vertex, couplings in the range from
$\alpha_0=0.1$ to 0.6 were used, while couplings up to $\alpha_0=1.75$
were used with both the bare and the Curtis-Pennington vertices.
Various renormalization points and renormalized masses were studied.

The subtractive renormalization procedure is straightforward to
implement.  The solutions are stable
and the renormalized quantities become independent of regularization
as the regularization is removed, which is as expected.
For example, the mass function $M(p^2)$ and momentum-dependent renormalization
$Z(p^2)$ are unchanged
to within the numerical accuracy of the computation as the
integration cutoff is increased by many orders of magnitude.
For the range of couplings $\alpha_0$ considered, the values of
the renormalization constant $Z_1(\mu,\Lambda) = Z_2(\mu,\Lambda)$
are never very far from 1 and vary relatively weakly with the choice of
renormalized mass and cutoff as expected in Landau gauge.  The mass
renormalization constant $Z_m(\mu,\Lambda)$ converges with increasing
$\Lambda$ because the mass function
$M(p^2)$ falls to zero sufficiently rapidly at large $p^2$.
This is a purely nonperturbative result and is in sharp contrast to the
perturbative case where this constant diverges.
For subcritical couplings, using the Curtis-Pennington vertex,
we also find that the mass renormalization
$Z_m(\mu,\Lambda)$ scales approximately as
$Z_m(\mu,\Lambda) \propto
    (\mu^2/\Lambda^2)^{(\frac{1}{2} - \gamma_{\rm CP}(\alpha_0))}$,
where e.g. $\gamma_{\rm CP}(0.5)= 0.358$.

Extensions of this work to include other gauges, other regularization
schemes (e.g., soft cut-offs), and vertices of the
Bashir-Pennington type are underway \cite{new-work}.

\begin{acknowledgements}

This work was partially supported by the Australian Research Council,
by the U.S. Department of Energy
through Contract No. DE-FG05-86ER40273, and by the Florida State University
Supercomputer Computations Research Institute which is partially funded by
the Department of Energy through Contract No. DE-FC05-85ER250000.
This research was also partly supported by grants of
supercomputer time from the U.S. National Energy Research Supercomputer
Center and the Australian National University
Supercomputer Facility.

\end{acknowledgements}

%=======================================================================
%          Bibliography:
%-----------------------------------------------------------------------

%=======================================================================
%        figures:
%-----------------------------------------------------------------------

\begin{figure}[tb]
  \centering
  %%\epsffile{CP_allAs.eps}
  %%\vspace{0.25in}
  %%\epsffile{CP_allMs.eps}
  \parbox{130mm}{\caption{Family of DSE solutions with the
    Curtis-Pennington vertex and subcritical couplings $\alpha_0 = 0.1$
    to 0.9.  The renormalization point is $\mu^2 = 100$, and
    renormalized mass is $m(\mu)=10$ for all cases shown.
    Ultraviolet cutoffs are $1 \times 10^{12}$ for all cases.
    (a) the finite renormalizations $A(p^2)$;
    (b) the mass functions $M(p^2)$.
  \label{fig_family} }}
  \vspace{0.25in}
\end{figure}

\begin{figure}[tb]
  \centering
  %%\epsffile{compare_As.eps}
  %%\vspace{0.25in}
  %%\epsffile{compare_Bs.eps}
  \parbox{130mm}{\caption{Comparison of the solutions to the DSE with the
    three different vertices.  All solutions have the subcritical coupling
    $\alpha_0 = 0.5$, renormalization point $\mu^2 = 100$, and
    renormalized mass $m(\mu)=10$.
    Ultraviolet cutoffs are $1 \times 10^{12}$ for all cases.
    (a) the finite renormalizations $A(p^2)$;
    (b) the scalar self-energy $B(p^2)$.
  \label{fig_compare3} }}
  \vspace{0.25in}
\end{figure}

\begin{figure}[tb]
  \centering
  %%\epsffile{scaled_Ms.eps}
  \parbox{130mm}{\caption{Asymptotic scaling of the mass function for
    the three vertex {\it Ans\"{a}tze\/} at
    the subcritical coupling $\alpha_0 = 0.5$; the
    functions plotted are $M(p^2) \times (p^2)^{\frac{1}{2}-\gamma'}$,
    where $\gamma' = .3614329$ is the correct anomalous dimension for
    the bare-vertex solution at this coupling.  The anomalous dimensions
    extracted for the Ball-Chiu and Curtis-Pennington vertices
    respectively are 0.35836  and  0.35843.
  \label{scaled_Ms} }}
  \vspace{0.25in}
\end{figure}

\begin{figure}[tb]
  \centering
  %%\epsffile{Zm_scaling.eps}
  \parbox{130mm}{\caption{The mass renormalization $Z_m$
    as a function of $\mu^2/\Lambda^2$, for the subcritical
    coupling $\alpha_0 = 0.5$.
    Bare masses for all solutions were chosen
    to give mass $m(p^2) = 10$ at $p^2 = 100$.
    Boxes, $\Box$, connected by solid lines, are results with cutoff
    $\Lambda^2 = 1 \times 10^{12}$; pluses, $+$, connected by dashed
    lines, are results with $\Lambda^2 = 1 \times 10^{18}$.
    Data are given in Table \protect{\ref{table_Zs}}.
  \label{fig_Zm_scaling} }}
  \vspace{0.25in}
\end{figure}

%%%\begin{figure}[tb]
%%%  %%  \setlength{\epsfxsize}{5.5in}
%%%  \centering
%%%  %%\epsffile{cutoff_indep.ps}
%%%  \parbox{130mm}{\caption{Infrared behavior of the dynamical mass
%%%    with the Curtis-Pennington vertex and subcritical coupling
%%%    $\alpha_0 = 0.5$, as the cutoff is varied over six orders of
%%%    magnitude.  Ultraviolet cutoffs for the two solutions are
%%%    $1 \times 10^{12}$ and $1 \times 10^{18}$.  The solutions are
%%%    equal to within their numerical accuracy.
%%%  \label{cutoff_indep} }}
%%%  \vspace{0.25in}
%%%\end{figure}

%%%\begin{figure}[tb]
%%%  %%\setlength{\epsfxsize}{5.5in}
%%%  \centering
%%%  %%\epsffile{Ren_A_1.eps}
%%%  \parbox{130mm}{\caption{The fermion function $A(p^2)$ for supercritical
%%%    coupling $\alpha_0=1.00$ for the Curtis-Pennington vertex.
%%%    Renormalization point is $\mu^2 = 100$,
%%%    with renormalized mass $m(\mu) = 10$.  The cutoff is stepped by
%%%    factors of 100 over eight orders of magnitude.  All five
%%%    solutions coincide to within the accuracy of the calculation
%%%    and are indistinguishable in this figure.
%%%  \label{Ren_A} }}
%%%  \vspace{0.25in}
%%%\end{figure}

\begin{figure}[tb]
  \centering
  %%\epsffile{Ren_B_1.eps}
  \parbox{130mm}{\caption{The scalar self-energy $B(p^2)$, for
  supercritical coupling $\alpha_0 = 1.00$ for the Curtis-Pennington
  vertex.  Renormalization point is $\mu^2 = 100$, with
  renormalized mass $m(\mu) = 10$.  The cutoff is stepped by
  factors of 100 over eight orders of magnitude.  Note that the
  mass curves all coincide.  Above $p^2 = 2\times 10^{11}$, the
  mass curves are negative, but the maximum negative excursion
  is $-2.4 \times 10^{-4}$.
  \label{Ren_B} }}
  \vspace{0.25in}
\end{figure}

\begin{figure}[tb]
  \centering
  %%\epsffile{CPvsbare.eps}
  \parbox{130mm}{\caption{Mass functions for the supercritical coupling
    $\alpha_0 = 1.15$ with
    both the bare and Curtis-Pennington vertices.  The renormalization
    point is $\mu^2 = 1 \times 10^8$, and $m(\mu) = 400$.
  \label{fig_CPvsbare} }}
  \vspace{0.25in}
\end{figure}

\begin{figure}[tb]
  \centering
  %%\epsffile{zcross_24.eps}
  \parbox{130mm}{\caption{Zero-crossings in the self-energy $B(p^2)$.
    The case shown has $\alpha_0=1.75$, $\mu^2 = 1 \times 10^4$,
    and $m(\mu) = 500$.  Note that there are {\it two\/}
    zero-crossings, and that the function approaches zero from
    above as $p^2 \to \infty$.  Note also that the negative peak
    is typically small compared with $B(0)$, $m(\mu)$, and $p^2$.
  \label{fig_zcross} }}
  \vspace{0.25in}
\end{figure}

%=======================================================================
%           Tables:
%-----------------------------------------------------------------------
\begin{table}
  \caption{Renormalization constants $Z_2(\mu,\Lambda)$, bare masses
    $m_0(\Lambda)$, and mass renormalization $Z_m(\mu,\Lambda)$ for
    various subcritical couplings for the Curtis-Pennington vertex,
    all with $\mu^2 = 100$, $m(\mu) = 10$.
    Cutoffs $\Lambda^2$ are $1\times 10^{12}$ and $1\times 10^{18}$.}
%%  \setdec 0.0000
  \begin{tabular}{lllll}
    $\alpha_0$ & $\Lambda^2$ & $Z_2(\mu,\Lambda)$ & $m_0(\Lambda)$ &
         $Z_m(\mu,\Lambda)$\\
    \tableline
      0.1 & $1 \times 10^{12}$ & 0.99822798  & 5.74672  & 0.574672 \\
	  & $1 \times 10^{18}$ & 0.99982280 & 4.0972   & 0.40972  \\
    \tableline
      0.2 & $1\times 10^{12}$ & 0.99931590 & 3.20125  & 0.320125 \\
	  & $1 \times 10^{18}$ & 0.99931590 & 1.59519  & 0.159519 \\
    \tableline
      0.3 & $1 \times 10^{12}$ & 0.99851308 & 1.7152   & 0.17152  \\
	  & $1 \times 10^{18}$ & 0.99851308 & 0.58245  &
	      $5.8245  \times 10^{-2}$ \\
    \tableline
      0.4 & $1 \times 10^{12}$ & 0.99744429 & 0.874486 &
	      $8.74486 \times 10^{-2}$ \\
	  & $1 \times 10^{18}$ & 0.99744428 & 0.195803 &
	      $1.95803 \times 10^{-2}$ \\
    \tableline
      0.5 & $1 \times 10^{12}$ & 0.99613623 & 0.417692 &
	      $4.17692 \times 10^{-2}$ \\
	  & $1 \times 10^{18}$ & 0.99613623 & 0.0590025 &
	      $5.90025 \times 10^{-3}$ \\
    \tableline
      0.6 & $1 \times 10^{12}$ & 0.99461286 & 0.18240  &
	      $1.8240 \times 10^{-2}$ \\
	  & $1 \times 10^{18}$ & 0.99461286 & 0.015271 &
	      $1.5271 \times 10^{-3}$  \\
    \tableline
      0.7 & $1 \times 10^{12}$ & 0.99289572 & 0.069783 &
	      $6.9783 \times 10^{-3}$  \\
	  & $1 \times 10^{18}$ & 0.99289572 & $3.14581 \times 10^{-3}$ &
	      $3.14581 \times 10^{-4}$ \\
    \tableline
      0.8 & $1 \times 10^{12}$ & 0.99100430 & 0.0214029 &
	      $2.14029 \times 10^{-3}$ \\
	  & $1 \times 10^{18}$ & 0.99100429 & $4.37117 \times 10^{-4}$ &
	      $4.37117 \times 10^{-5}$ \\
    \tableline
      0.9 & $1 \times 10^{12}$ & 0.98895625 & $4.01753 \times 10^{-3}$ &
              $4.01753 \times 10^{-3}$ \\
	  & $1 \times 10^{18}$ & 0.98895626 & $2.31378 \times 10^{-5}$ &
              $2.31378\times 10^{-5}$ \\
  \end{tabular}
  \label{table_Zs_by_a}
\end{table}

\begin{table}
  \caption{Renormalization constants $Z_2$ and $Z_m$ for several
    different ratios $\mu^2/\Lambda^2$; all solutions are for the subcritical
    coupling $\alpha_0 = 0.5$ for the Curtis-Pennington vertex, with
    renormalized masses chosen as $m(\mu=10) = 10$.}
%%  \setdec 0.0000
  \begin{tabular}{lllll}
    & \multicolumn{2}{c}{$\Lambda^2 = 1 \times 10^{12}$}
    & \multicolumn{2}{c}{$\Lambda^2 = 1 \times 10^{18}$} \\
    $\mu^{2}/\Lambda^{2}$ & $Z_2$ & $Z_m$ & $Z_2$ & $Z_m$\\
    \tableline
     $1\times 10^{-10}$ & 0.99613623     & 0.04177
         & 0.999999999439219  &  0.032364 \\
     $1\times 10^{-9}$  & 0.99914330     & 0.04836
         & 0.999999999970789  &  0.044846 \\
     $1\times 10^{-8}$  & 0.99993219     & 0.06319
         & 0.999999999998478  &  0.06214 \\
     $1\times 10^{-7}$  & 0.99999612     & 0.08640
         & 0.999999999999921  &  0.08611 \\
     $1\times 10^{-6}$  & 0.99999979     & 0.11940
         & 0.999999999999995892 & 0.11932 \\
     $1\times 10^{-5}$  & 0.99999998     & 0.16536
         & 0.999999999999999778 & 0.16534 \\
     $1\times 10^{-4}$  & 0.9999999994   & 0.22912
         & 1.000000000000000  &   0.22911 \\
     $1\times 10^{-3}$  & 0.99999999997  & 0.31749 \\
  \end{tabular}
  \label{table_Zs}
\end{table}

\begin{table}
  \caption{Renormalization constants $Z_2(\mu,\Lambda)$, bare masses
    $m_0(\mu,\Lambda)$, and mass renormalizations $Z_m(\mu,\Lambda)$
    for the supercritical coupling $\alpha_0 = 1.15$ for the Curtis-Pennington
    vertex.  All solutions are renormalized
    with $\mu^2 = 1 \times 10^8$, $m(\mu) = 400$;
    the cutoff is stepped from $1 \times 10^8$ to $1 \times 10^{12}$.}
  \begin{tabular}{lllll}
    $\Lambda^2$ & $\mu^2/\Lambda^2$ & $Z_2$ & $m_0(\Lambda)$ & $Z_m$ \\
  \tableline
    $1 \times 10^{8}$  & $1$     & 0.999913461 &
	230.581 & 0.57645 \\
    $1 \times 10^{9}$  & $.1$    & 0.999848292 &
	53.5732 & 0.13393 \\
    $1 \times 10^{10}$ & $.01$   & 0.999846784 &
        4.44299 & $1.1107 \times 10^{-2}$ \\
    $1 \times 10^{11}$ & $.001$  & 0.999846902 &
	-3.93251 & $-9.8313 \times 10^{-3}$ \\
    $1 \times 10^{12}$ & $.0001$ & 0.999846922 &
	-2.84691 & $-7.1173 \times 10^{-3}$ \\
  \end{tabular}
  \label{table_Zs_1.15}
\end{table}

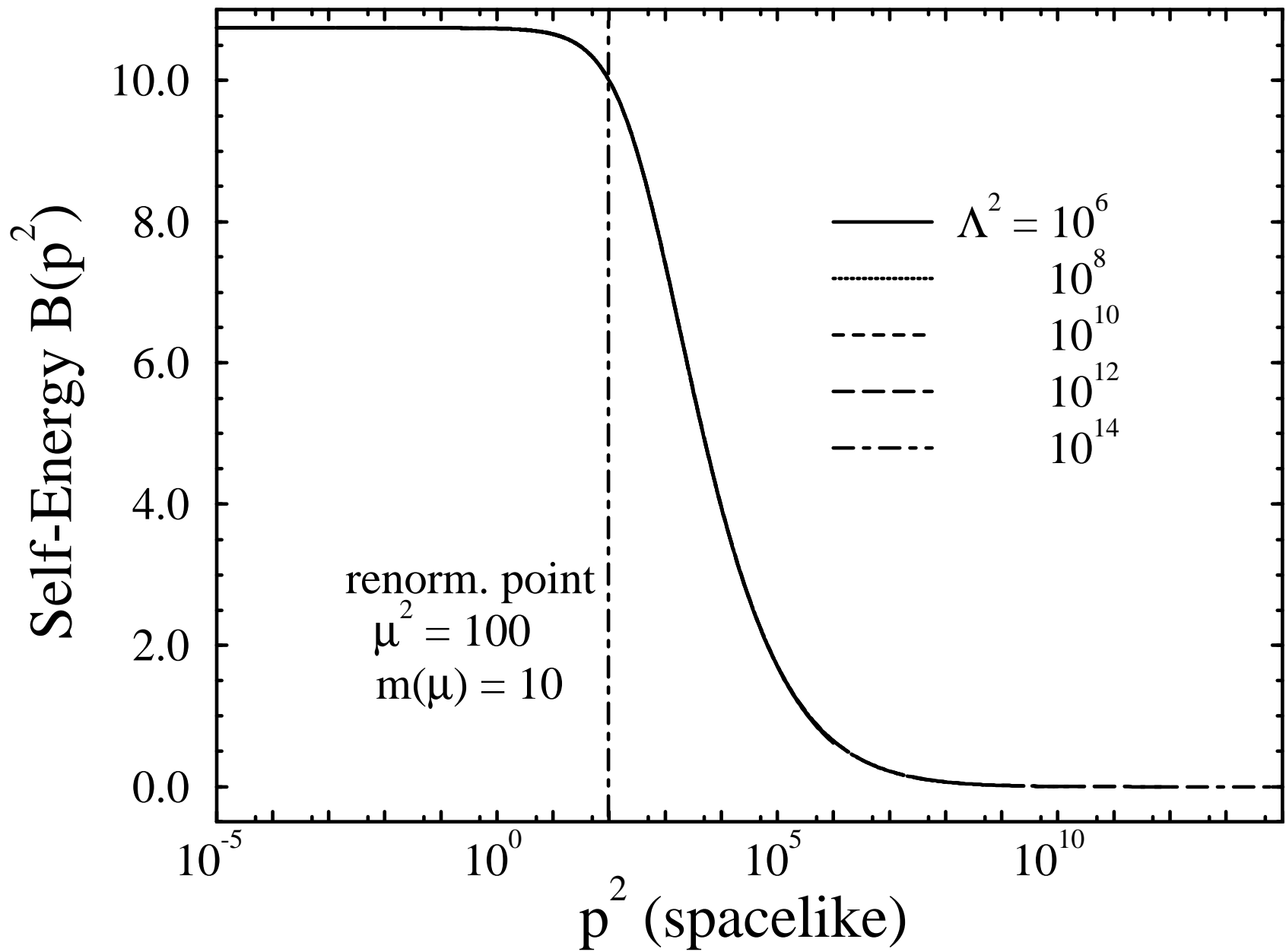
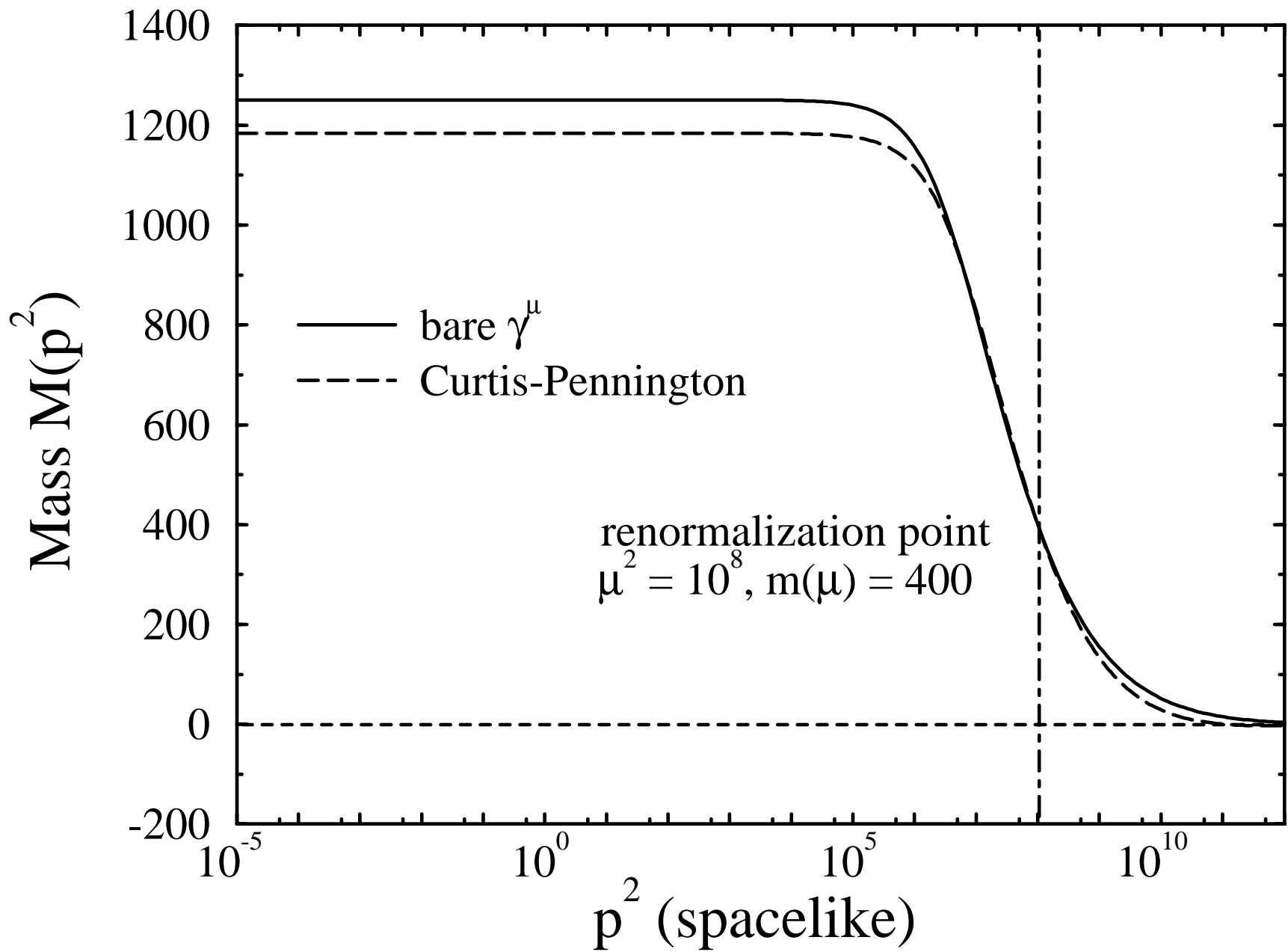
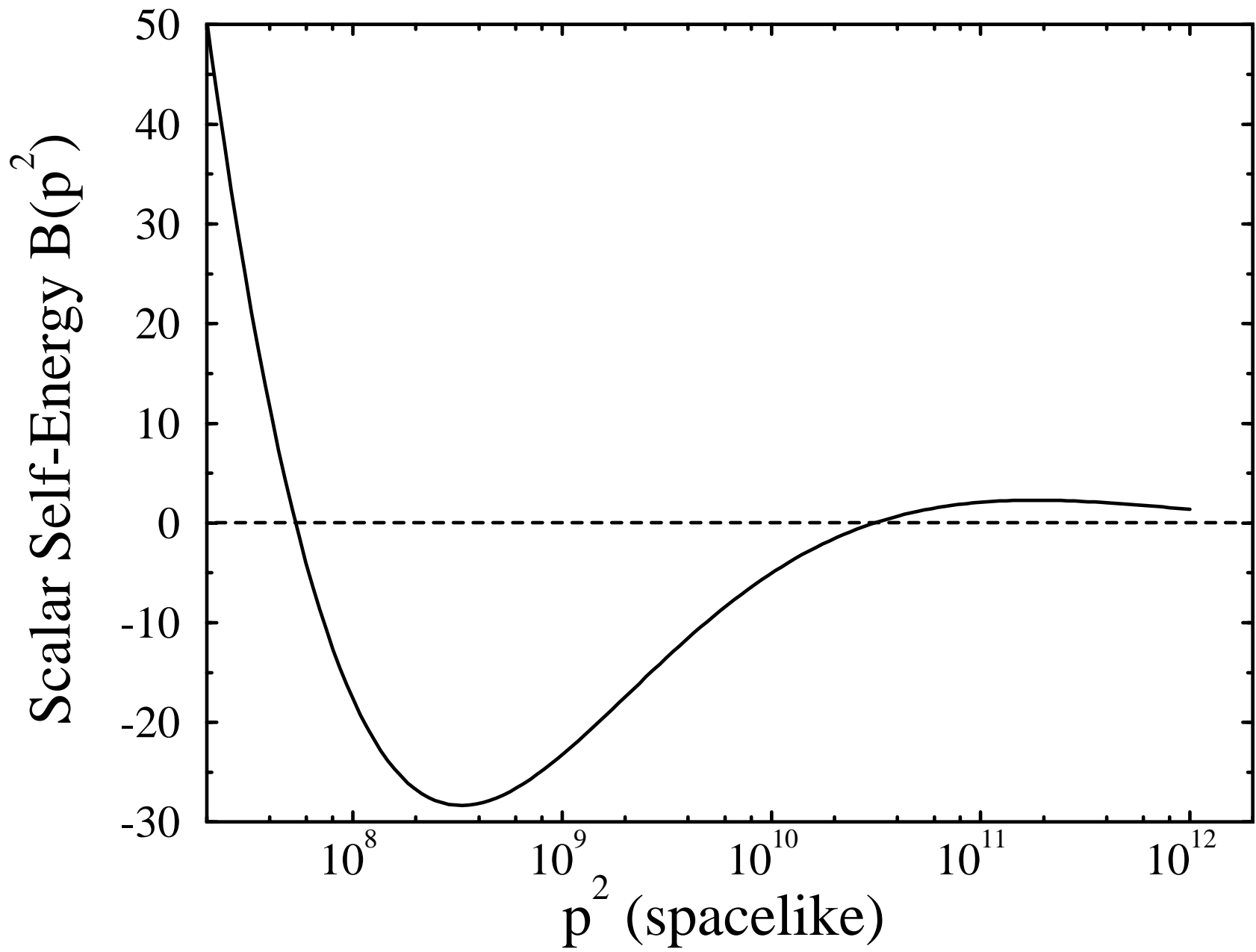
\end{document}